\documentclass[preprint,aps,prc,amsmath,amssymb,showpacs,showkeys,floatfix,diagbox]{revtex4-2}
\usepackage{graphicx}
\usepackage{dcolumn}
\usepackage{bm}
\usepackage{array,tabularx}
\usepackage{multirow}
\usepackage{xcolor}
\usepackage[flushleft]{threeparttable}
\usepackage{hyperref}
\begin{document}

\title{Alpha decay of thermally excited nuclei}
\author{J. E. Perez Velasquez}
\email{je.perez43@uniandes.edu.co}
\affiliation{Departamento de Fisica, Universidad de los Andes,
Cra.1E No.18A-10, Bogot\'a, Colombia}

\author{O. L. Caballero}
\email{ocaballe@uoguelph.ca}
\affiliation{Department of Physics, University of Guelph, Guelph, 
ON N1G 2W1, Canada}

\author{N. G. Kelkar}
\email{nkelkar@uniandes.edu.co}
\affiliation{Departamento de Fisica, Universidad de los Andes,
Cra.1E No.18A-10, Bogot\'a, Colombia}
\date{\today}

\begin{abstract}
One of the prominent decay modes of heavy nuclei which are produced in 
astrophysical environments at temperatures of the order of $10^9$ K 
is the $\alpha$ ($^4$He) decay. 
Thermally enhanced $\alpha$ decay rates are evaluated within the 
standard scheme of a tunneling decay where the $\alpha$ particle tunnels 
through the potential barrier formed by its interaction with the daughter 
nucleus. Following the observation that there exist several 
excited levels with the possibility of an $\alpha$ decay when the 
daughter nucleus is at a shell closure, we focus in particular on 
decays producing daughter nuclei with the neutron number, $N = 126$. 
Within a statistical approach we find that 
the half-lives, $t_{1/2}(T)$, for temperatures ranging from $T$ = 
0 to 2.4 GK can decrease by 1 - 2 orders of magnitude with the 
exception of the decay of $^{212}$Po which decays to the 
doubly magic daughter $^{208}$Pb, where $t_{1/2}(T)$ decreases by 5 orders
of magnitude. The effect of these thermally enhanced $\alpha$ decays on the 
$r$-process nucleosynthesis can be significant in view of the mass build 
up at the waiting point nuclei with closed neutron shells. 
  
\end{abstract}
\maketitle

\section{Introduction}

The unprecedented observational data from compact object mergers in the recent years have confirmed the general model of the synthesis of heavy elements via rapid-neutron captures \cite{Abbott2017a,Abbott2017b,kilonova-data,Chornock2017}. However, details required for the interpretation of the correlated ultraviolet, optical and infrared electromagnetic emissions, also known as kilonova (for a review see \cite{Metzger}), are still unclear. For example, accurate kilonova modeling  relies on the knowledge of the fraction of mass ejected during the event, its opacity, and the radioactive decay of freshly produced lanthanide and actinide nuclei \cite{Kasen-nature,Kasen-Barnes}.  

Nucleosynthesis involving the 
production of heavy and super heavy nuclei is expected to take place at 
very high temperatures of the order of a few Giga Kelvin (GK). The 
{\it transuranic}  
elements in particular are considered to be produced in the $r$-process
occuring in environments where the neutron captures are faster than the 
$\beta$ decays. 
The conditions during the r-process, involving densities, 
$\rho(n) \sim 10^{20} cm^{-3}$ and temperatures $T \sim $ 10$^{9}$ K,  
are explosive conditions \cite{bertulanibook}. 
The possible sites for the r-process are considered 
to be \cite{bertulanibook} (a) neutronized atmosphere above the  
proto-neutron star in
a Type II supernova, (b) neutron-rich jets from supernovae or neutron
star mergers, (c) inhomogeneous Big Bang, (d) He/C zones in Type II
supernovae, (e) red giant He flash, (f)  spallation neutrons in He
zone (g) neutrino driven wind from freshly born neutron stars and 
(h) outflows from black hole accretion discs originated in compact object mergers or collapsars (for recent reviews of the possible r-process sites see Cowan et al \cite{Cowan2021-sites} and C{\^o}t{\'e} et al \cite{Cote2019}).  
The abundance of elements is found through a network \cite{network,Skynet} 
of coupled differential equations involving nuclear reaction rates at elevated 
temperatures. Theory and models play an important role in determining the 
latter for neutron-rich nuclei which cannot be measured in terrestrial laboratories.

The $r$-process nucleosynthesis path is along highly unstable, 
exotic, and neutron-rich nuclei that in principle does not involve alpha 
emitters. However, once heavy neutron rich nuclei in the region with Z $>$ 82 
are formed, and with the depletion of further neutron captures
(i.e. after r-process freeze out), those nuclei decay by different modes
(e.g. beta, alpha, fission). Some decay modes would compete. This
process can even lead to the formation of nuclei in the actinide region.
The nuclei studied in this work are part of the mass region (A $>$ 208) where
several alpha emitters are found \cite{martin,zhao}.
Thus, it is not only the photo-dissociation 
and neutron capture cross sections but also fission (spontaneous and induced) 
and the decay rates which are important 
for the abundance evolution. The explosive conditions in supernovae and neutron star mergers \cite{thieleman,arcones,albino} leading to considerably high temperatures could result in nuclei existing in excited states. Though the possible influence of these nuclear thermal excitations 
is taken into account in the production reactions as well as in 
their reverse reactions, with libraries publicly available for the scientific community (e.g. \cite{Reaclib,Starlib,bruslib}), the same is not true in the 
case of $\alpha$ decay. These decay rates, 
entering as an input to the network calculations, are taken to be the 
ground state (or terrestrial) half-lives \cite{network}. However, one must note that for high ambient temperatures, 
the population factor for the excited energy levels of a nucleus 
is large. These are thermal excitations and one must take into account the 
possibility of $\alpha$ decay of thermally excited nuclei. This is in particular quite important for the $r$-process nucleosynthesis
where the closed neutron shells present waiting points due to the fact that 
it takes a long time for the successive $\beta$ decays, which are slow at the shell closures, to occur 
and allow progression through higher N nuclei. In section \ref{structure}, we will see that 
the $\alpha$ decays of parent nuclei producing daughter nuclei at the shell 
closures display a stronger temperature dependence with increased decay 
rates at higher temperatures. 

Apart from the paper of Perrone and Clayton \cite{clayton}, published 
in 1970, there is indeed no estimate of the possible effects of temperature on the 
$\alpha$ decay half-lives. However, given the fact that about 50 years ago, 
the data on excited levels of nuclei was scarce, calculations were 
performed assuming a continuum of states described by the available density 
of states. The latter assumption as we will see leads to a very large 
overestimate of the enhancement in the decay rate due to temperature. 

In the present work, we investigate the temperature dependence of the 
$\alpha$ decay rates relevant for the $r$-process nucleosynthesis. The 
calculations are performed within two different approaches: (i) a statistical 
model which makes use of the experimentally measured excited levels and 
an empirical decay law (fitted to data) in the absence of available data 
on the half-lives and (ii) 
a theoretical model which treats the $\alpha$ decay as a semiclassical 
tunneling of the $\alpha$ particle through the barrier created by the 
interaction of the $\alpha$ and the daughter nucleus which exist inside 
the parent in the form of a preformed cluster. The latter calculation is 
performed using a density dependent folding model which has been reasonably 
successful in reproducing the measured $\alpha$ decay half-lives \cite{kelkar2,JhoanPRC}. The formalism is presented in sections \ref{formalism}, 
and \ref{T-dependence}. In section \ref{structure} we connect to 
shell closures, and in section \ref{results} we discuss the results. 
Finally, we summarize our findings in section \ref{conclusions}. 

\section{Alpha decay formalism}
\label{formalism}
One of the most successful achievements of the quantum theory is the 
explanation of the $\alpha$-decay of radioactive nuclei as a tunneling problem. 
This approach was developed independently by Gamow \cite{gamow} and by 
Gurney and Condon \cite{condon} in the late twenties.
Though $\alpha$-decay has been studied since then within different 
quantum mechanical approaches 
\cite{renreview}, semiclassical approaches based on the tunneling of an 
$\alpha$ particle through the potential barrier created by its interaction 
with the daughter nucleus produced in the decay are some of the most popular 
and widely used methods for calculating half-lives. 
The interaction potential between the $\alpha$ ($^4$He nucleus) and the 
daughter nucleus, and the $Q$-value, which is usually taken to be the 
energy of the tunneling $\alpha$, play the main role in determining the 
tunneling probability and hence the half-life.  
Using the JWKB approximation \cite{froman1}, 
different semiclassical approaches lead to the same expression for the $\alpha$-decay width \cite{kelkar2}
\begin{equation}
\Gamma=P_{\alpha}\frac{\hbar^2}{2\mu}\left[\int_{r1}^{r2}{\frac{dr}{\kappa(r)}}\right]^{-1}\exp\left[-2\int_{r2}^{r3}{\kappa(r)dr}\right]\label{width}
\end{equation}
with the so-called wave number $\kappa(r)=\sqrt{\frac{2\mu}{\hbar^2}|V(r)-Q|}$ 
and $\mu$ the reduced mass of the daughter-$\alpha$ system. 
The classical turning points $r_1$, $r_2$ and $r_3$ are obtained by solving 
the equation $V(r)= Q$ where $Q$ is the energy of the tunneling  
$\alpha$-particle. The factor in front of the exponential arises due to 
the normalization of the bound state wave function in the region between 
$r_1$ and $r_2$. The exponential factor is the penetration probability. 
The $\alpha$-decay half-life of an isotope is evaluated as
\begin{equation}
t_{1/2}=\frac{\hbar \ln 2}{\Gamma}\, .\label{halflife}
\end{equation} 
Since the tunneling decay assumes the existence of a preformed cluster of the 
$^4$He and daughter nucleus inside the decaying parent nucleus, one must 
include a preformation probability $P_{\alpha}$ in the expression for 
half-life. This factor, in principle, can be expressed as an overlap between 
the wave functions of the parent nucleus and the decaying-state wave function 
describing the $\alpha$-particle coupled to the daughter nucleus. 
Such a microscopic undertaking is still considered a difficult task and a 
phenomenological way to determine $P_{\alpha}$ is simply taking the ratio 
of the theoretical and experimental half-lives, such that, 
$P_{\alpha}={t_{1/2}^{theory}}/{t_{1/2}^{exp}}$, where, $t_{1/2}^{theory}$ 
is evaluated using $\Gamma$ from Eq. (\ref{width}) but with $P_{\alpha} = 1$.  

The total potential between the $\alpha$ and the daughter nucleus is 
typically written as a function of the distance between their centers of mass 
as, 
\begin{equation}
V(r)=V_n(r)+V_C(r)+\frac{\hbar^2 (l+1/2)^2}{2 \mu r^2}\label{potential}
\end{equation}
where $V_n(r)$ and $V_C(r)$ are the nuclear and Coulomb potentials, 
respectively. 
The last term in equation \ref{potential} represents the Langer 
modified centrifugal potential \cite{langer} which must be used while using 
the JWKB approximation. Some of the calculations presented in this work will 
be performed within the density dependent double folding model (DFM) which 
is based on realistic nucleon-nucleon interactions 
and has been reasonably successful in 
reproducing the experimental half-lives. The details of this potential can be found in \cite{JhoanPRC,DiegoJhoan2022}. Here we describe it briefly. In the DFM, the 
nucleus-nucleus interaction is related to the NN interaction by folding an effective NN interaction over the density distribution of the two nuclei. The folded nuclear potential is written as
\begin{equation}\label{eq:V_Double-Folding}
    V_n(\mathbf{r})= \lambda \int \mathrm{d} \mathbf{r}_{1} \mathrm{d} \mathbf{r}_{2} \,\rho_{\alpha}\left(\mathbf{r}_{1}\right)v_N\left(|\mathbf{s}|=\left|\mathbf{r}+ \mathbf{r}_{2}-\mathbf{r}_{1}\right|\right) \rho_{d}\left(\mathbf{r}_{2}\right)\,.
\end{equation}
where $\rho_i$ ($i=$ $d$, $\alpha$) are the densities of the alpha and the daughter nucleus in a decay, and $v_N(|\mathbf{s}|)$ is the nucleon-nucleon (NN) interaction (see \cite{DiegoJhoan2022} for the figure with details). The matter density distribution of the heavy daughter is calculated as
\begin{equation}
    \rho(r)=\frac{\rho_{0}}{1+\exp \left(\frac{r-R}{a}\right)}\,,
\end{equation}
where $\rho_0$ is obtained by normalizing $\rho(r)$ to the mass number, $\int\rho(\mathbf{r})\,{\rm d}\mathbf{r}=A$, and the constants are given as $R=1.07A^{1/3}$ fm and $a=0.54$ fm. The alpha or $^4$He density distribution is given using a standard Gaussian form \cite{satchler}, namely,
\begin{equation}
    \rho_{\alpha}(r) = 0.4229\exp\left(-0.7024\,r^2\right)\,.
\end{equation}

We use the popular choice of the effective NN interaction which is based on the M3Y-Reid-type soft core potential,
\begin{equation}
    v_N\left(|\mathbf{s}|\right)=7999 \frac{\exp \left(-4\left|\mathbf{s}\right|\right)}{4\left|\mathbf{s}\right|}-2134 \frac{\exp \left(-2.5\left|\mathbf{s}\right|\right)}{2.5\left|\mathbf{s}\right|}\, + \,J_{00}\delta(\mathbf{s}),
\end{equation}
where $|\mathbf{s}|=\left|\mathbf{r}+\mathbf{r}_{2}-\mathbf{r}_{1}\right|$ is the distance between a nucleon in the daughter nucleus and a nucleon in the alpha. The above NN interaction consists of a short-ranged repulsive part and a long-ranged attractive one, in addition to the zero-range contribution $J_{00}\delta(\mathbf{s})$ with $J_{00}=-276(1-0.005\,E/A_c)$. The latter is the so-called knock-on exchange term which takes into account the antisymmetrization of identical nucleons in the alpha and the daughter nucleus. It represents a kind of nonlocality in the DFM potential and in order to avoid double counting, is usually not included in the calculation if one uses nonlocal nuclear potentials \cite{DiegoJhoan2022}. The strength of the nuclear potential, $\lambda$, is deduced by requiring the Bohr-Sommerfeld quantization condition to be satisfied \cite{JhoanPRC}. The Coulomb potential, $V_C(r)$, is obtained in a similar way with the matter densities of the alpha and the daughter replaced by their charge densities (which have the same form as above but are normalized to the number of protons).   

The angular momentum, $l$, 
carried by the alpha particle must satisfy the following spin-parity 
selection rules,
\begin{equation}
|J_p-J_d| \leq l \leq |J_p+J_d|\quad\mathrm{and}\quad \pi_d=\pi_p(-1)^l \label{selectionrules}
\end{equation}
where ($J_p$, $\pi_p$) and ($J_d$, $\pi_d$) are the (spin, parity) of the 
parent and daughter nuclei, respectively. 

In Table \ref{halflivesdoublefoldingmodel} we present the half-lives for 
some nuclei using the DFM. 
We examine transitions for which the alpha particle has the minimum 
angular momentum value, $l_{min}$, 
satisfying equations \eqref{selectionrules}. For the 
decays considered in Table \ref{halflivesdoublefoldingmodel}, $l_{min} = 0$.  
The experimental half-lives and the corresponding 
preformation factors are also listed in Table \ref{halflivesdoublefoldingmodel}.
The theoretical values obtained are close to some others found in the 
literature \cite{bairenroepke}.  
\begin{table}[h]
\renewcommand{\arraystretch}{1.5}
\resizebox{\textwidth}{!}{
\centering
\begin{tabular}{|p{2.1cm}|p{2.1cm}|p{2.1cm}|p{2.5cm}|p{2cm}|p{2.5cm}|}\hline
 Isotope &Q-Value & $t_{1/2}^{exp}$ & $t_{1/2}^{\rm[DFM]}$ & 
$P_{\alpha}{\rm [DFM]}$&$t_{1/2}^{\rm [UDL]}$    \\
 &[MeV]   & [s]                & [s]                   &   & [s] \\
\hline
 $_{84}^{212}$Po & 8.954 & $2.9 \times 10^{-7}$  &$6.49 \times 10^{-8}$ & 0.22 & $ 1.57\times 10^{-7}$ \\
 $_{86}^{214}$Rn  & 9.208 & $2.7\times 10^{-7}$  & $7.91\times 10^{-8}$ & 0.29 & $2.06\times 10^{-7}$ \\
 $_{87}^{215}$Fr  & 9.540 & $8.6\times 10^{-8}$ & $2.9\times 10^{-8}$ & 0.33 & $7.10\times 10^{-8}$\\
  $_{88}^{216}$Ra   & 9.526 & $1.8\times 10^{-7}$  &  $6.86\times 10^{-8}$  & 0.37 & $1.86\times 10^{-7}$   \\
  $_{89}^{217}$Ac  & 9.832 & $6.9\times 10^{-8}$  & $2.98\times 10^{-8}$ & 0.43& $7.67\times 10^{-8}$  \\
 \hline
\end{tabular}}
\caption{Comparison of the alpha decay half-lives evaluated using the double 
folding model (DFM) and a universal decay law (UDL) (\ref{universallaw}), 
with experiments. 
The phenomenological preformation factors, 
$P_{\alpha} = {t_{1/2}^{\rm [DFM]}}/{t_{1/2}^{exp}}$ using the DFM 
half-lives are also shown.
Here, $t_{1/2}^{\rm [DFM]}$ 
is evaluated using $\Gamma$ from Eq. (\ref{width}) but with $P_{\alpha} = 1$
}\label{halflivesdoublefoldingmodel}
\end{table}

The double folding model calculations can in principle be improved with the 
inclusion of deformation and nonlocalities in the interaction potential 
\cite{JhoanPRC,DiegoJhoan2022}. 
However, the objective of the present work is to 
perform a comparative study of approaches for  
half-lives measured on earth and in a hot astrophysical environment and
hence it suffices to perform calculations within a model which can reproduce 
alpha decay half-lives reasonably well.
Indeed, we shall also use an empirical formula (a universal decay law (UDL) 
for $\alpha$ and cluster decay, obtained from fits to extensive data) 
for the half-lives 
calculated within the statistical approach (to be explained in the next 
section) since (i) the UDL gives the right order 
of magnitude estimate of half-lives and (ii) it would be quite a tedious 
undertaking to evaluate the half-lives of several excited states within the 
double folding model (DFM) without any significant advantage.
Such universal decay laws are usually obtained by starting with an 
analytical expression \cite{BookBeisser} 
which is based on the assumption of a rectangular 
well for the nuclear potential and a point-like Coulomb potential between the 
decay products of the radioactive nucleus. The constants appearing in such 
an expression are then assumed to be free parameters and fitted to an 
extensive set of data. 
The latter compensates for the simplistic assumptions made 
in the derivation of the empirical formula and provides a useful expression 
depending on the number of nucleons and the $Q$-value of the decay. 
We use the following UDL obtained in \cite{Qi,QiPRL}.
\begin{equation}\label{universallaw}
\log_{10} t_{1/2}=2aZ_d\sqrt{\frac{A}{Q}}+b\sqrt{4AZ_d(A_d^{1/3}+4^{1/3})}+c+d\sqrt{4AZ_d(A_d^{1/3}+4^{1/3})}\sqrt{l(l+1)}
\end{equation}
where $A=\frac{4A_d}{A_d+4}$, with $A_d$, $Z_d$ the mass and proton 
numbers of the daughter nucleus respectively, and the constants are 
given by $a=0.4392060$, $b=-0.3944174$, $c=-27.0648730$ and $d=0.0051825$ such 
that $Q$ is taken in MeV resulting in the half-life, $t_{1/2}$, in seconds.

In the next section we shall also describe the effective $Q$-value approach 
to evaluate the temperature dependent half-lives using the DFM and the UDL.  
Comparing it with the statistical approach gives us an idea of the usefulness 
of this approach in the context of $\alpha$ 
decay half-lives in an astrophysical 
environment. We shall also discuss one of the earliest attempts to evaluate 
the thermally enhanced $\alpha$-decay rates in connection with the $s$-process 
nucleosynthesis \cite{clayton}. The authors in \cite{clayton} predicted a
decrease in the half-lives by about 20 - 60 orders of magnitude depending 
on the nucleus for a temperature around 2 GK. We do not find such spectacular 
effects of temperature in our calculations. The reason for this difference 
will become evident in the next sections.

\section{Temperature dependent half-lives}
\label{T-dependence}

Nucleosynthesis in the later stages of stellar evolution, especially 
through the r-process is considered to take place at considerably 
high temperatures of the order
of 10$^9$ K. The calculation of the abundance of heavy elements depends on a precise 
determination of the nuclear reaction rates of the processes which produce 
the elements as well as the processes which destroy the newly formed 
nuclei. 
Though the neutron capture cross sections and their reverse reaction rates 
at elevated temperatures are carefully taken into account, the network codes 
usually rely on the laboratory values of the half-lives of $\alpha$ 
decays from the ground states of nuclei. 
Perrone and Clayton \cite{clayton} investigated the effect of thermally excited states in the alpha decay of some nuclei and its application in the $s$-process nucleosynthesis. However, such effects are not included in the $r$-process 
simulations. 

In what follows, we present a statistical
calculation of thermally enhanced $\alpha$-decay rates 
that includes experimentally observed excited levels for some nuclei. 
We also propose a model that uses an effective $Q$-value approach and 
which makes use of an average excitation energy at a given temperature.  
Finally, we discuss the approach of Perrone and Clayton briefly for 
completeness. 

\subsection{Statistical calculation}
The temperature-dependent half-life, $t_{1/2}(T) = \ln 2/\lambda(T)$, 
can be evaluated within the standard statistical approach \cite{iliadisbook} 
by defining the temperature dependent decay constant as follows:  
\begin{equation}
\lambda(T)=\sum_ip_i\sum_{j}\lambda_{ij}\,.\label{Iliadis}
\end{equation}
Here the sums over $i$ and $j$ are over the parent and daughter states 
respectively. Thus, $\lambda_{ij}$ is the decay constant for 
the decay of the $i^{th}$ level in the parent to the $j^{th}$ level in 
the daughter such that 
\begin{equation}
\lambda_{ij}=\frac{\ln(2)}{t_{1/2}^{ij}}\,.
\end{equation}
The population probability, $p_i$, is given with a Boltzmann factor as 
\cite{WardFowler} 
\begin{equation}\label{populate}
p_i=\frac{(2J_i+1)e^{(-E_i/k_BT)}}{\sum_l(2J_l+1)e^{(-E_l/k_BT)}}
\end{equation}
where $J_i$ and $E_i$ are the spin and the excitation energy of the 
state $i$, respectively. 
Inserting (\ref{populate}) 
in (\ref{Iliadis})   
\begin{equation}\label{stathalflife}
\lambda(T)=\frac{\ln(2)}{\sum_l(2J_l+1)e^{(-E_l/k_BT)}}\sum_{i,j}{\frac{(2J_i+1)
e^{(-E_i/k_BT)}}{t_{1/2}^{i}}} \,(BR)_{ij}
\end{equation}
where $(BR)_{ij}$ is the branching fraction for the decay from the $i^{th}$ 
level of the parent nucleus to the $j^{th}$ level in the 
daughter nucleus. 
The detailed decay 
schemes and the percentage decay to a particular channel, i.e., 
$I = [\lambda_{ij}/\lambda_{tot}]*100\%$ can be found at the web-site in 
\cite{halflifedata}. The branching fraction, $(BR)_{ij} = 
\lambda_{ij}/\lambda_{tot}$, can thus be obtained from the data tables.   
To evaluate the temperature dependent half-life in the statistical approach, 
we shall use Eq. (\ref{stathalflife}) with the input half-lives, 
$t_{1/2}^{i}$ and $(BR)_{ij}$ taken from experiment.  
If the experimental half-life of a level
is not known, it is calculated using the UDL at an effective $Q$-value given
by $Q + E_i$, where $E_i$ is the energy of the excited level. In such cases,
even if the experimental branching ratio is known, it is not used but taken to
be 100\% since the UDL per definition is formulated only for the alpha decay
channel. 

\subsection{Effective Q-value model}\label{Qeff-model}
Alpha decay half-lives are very sensitive to the penetrability factor which 
at the same time means that they are very sensitive to the $Q$-values too. 
For a tunneling decay of an $\alpha$ particle taking place in a very hot 
surrounding, one can model the effect of temperature by an increase in the 
energy, $Q$, with which the $\alpha$ tunnels through the Coulomb barrier. 
A higher $Q$-value would clearly reduce the area under the integral in the 
exponential term in Eq. (\ref{width}) and hence lead to an increase in 
the decay probability (and therefore a decrease in the half-life). 
This increase in the $Q$-value can be modelled by adding an average 
excitation energy of the nucleus at a given temperature.
Such an effective $Q$-value of an $\alpha$-decay can be expressed as
\begin{equation}
Q_{eff}=Q + \bar{\epsilon}(A,Z,T)\label{effectiveQ}
\end{equation}
where $\bar{\epsilon}(A,Z,T)$ is given by the standard definition 
of the average 
excitation energy \cite{sitenko} in statistical physics as,   
\begin{equation}
\bar{\epsilon}(A,Z,T)=-\frac{\partial}{\partial\beta}\ln {\sf Z}(A,Z,T)
\label{aee}
\end{equation}
with the canonical partition function ${\sf Z}$ given by
\begin{equation}
{\sf Z}(A,Z,T)=\sum_i^n g_i\exp(-\beta E_i)+\int_{E_n}^{E_{max}}{D(E)\exp(-\beta E)dE}
\end{equation}
where, $\beta = 1/(k_B T)$, $g_i=2J_i+1$ and $J_i$ is the spin of the 
$i^{th}$ level and 
$D(E)$ is the nuclear level density for which we choose the following form 
\cite{ericsonref}: 
\begin{equation}
D(E)=\frac{\sqrt{\pi}}{12}\frac{e^{2(aE)^{1/2}}}{a^{1/4}E^{5/4}}\label{ericson}
\end{equation}
The level density parameter, $a$, is taken to be $A/9$ where $A$ is 
the mass number of the parent nucleus.
If we consider the discrete levels as well as the continuum, the average 
excitation energy, using the above equations can be expressed as 
\begin{equation}
\bar{\epsilon}(A,Z,T)=\frac{\sum_i^n g_iE_i\exp(-\beta E_i)+\int_{E_n}^{E_{max}}{ E\times D(E)\exp(-\beta E)dE}}{\sum_i^n g_i\exp(-\beta E_i)+\int_{E_n}^{E_{max}}{D(E)\exp(-\beta E)dE}}.
\label{excitation2}
\end{equation}
The temperature, $T$, appearing in the above expression (through $\beta = 1/(k_B T)$), is in principle 
the {\it nuclear temperature}. Modification of nuclear properties at finite 
temperatures is relevant both for applications in astrophysics \cite{Lattimer,
BetheRevModPhys,Botvina} and for
models of finite nuclei and nuclear matter at high excitation
energy \cite{Benvenuto,Morrisey}.
The internal excitations of nuclei can  
play an important role in regulating their abundance.
The excitations form an important ingredient in multifragmentation studies
of hot nuclei. For example, the authors in \cite{BotvinaPLB}, working
within a statistical multifragmentation model find significant
temperature dependent modifications
relevant for stellar dynamics and nucleosynthesis.
They perform calculations
for supernova matter by assuming that the nuclei have the same internal
temperature as the surrounding medium. In the present work, we shall also 
assume a dynamical equilibrium such that the temperature, $T$, above can be 
assumed to be the surrounding temperature.

Having defined the effective $Q$-value in this manner, the temperature 
dependent half-life within the density dependent double folding model 
can be evaluated using Eq. (\ref{width}) with $Q$ replaced by $Q_{eff}$. 
With a similar replacement we can also estimate the temperature 
dependence using the UDL in 
Eq. (\ref{universallaw}). In Table \ref{table1} we list the average 
excitation energies at different temperatures for some heavy nuclei which will
be studied in this work. 
\begin{table}[h]
\resizebox{\textwidth}{!}{
\begin{tabular}{ |p{1.9cm}|p{1.9cm}|p{1.9cm}|p{1.9cm}|p{1.9cm}| p{1.9cm}|p{1.9cm}|}
 \hline
 \multicolumn{7}{|c|}{$\bar{\epsilon}(A,Z,T)$ [MeV]} \\
 \hline
&$Q$ \,[MeV]&0.8 GK&1.2 GK&1.6 GK& 2 GK&2.4 GK\\ \hline
$_{\,\,\,84}^{212}$Po &8.954 &$0.000096$  & $0.0034$ & $0.0231$ & $0.08142$ & $0.2004$\\

$^{214}_{\,\,\,86}$Rn &9.208 &$0.000147$  & $0.00438$ & $0.02520$ & $0.0754$ & $0.1609$\\

$^{215}_{\,\,\,87}$Fr &9.540 &$0.00096$  & $0.02738$ & $0.1349$ & $0.3031$ & $0.4646$\\

 $^{216}_{\,\,\,88}$Ra &9.526&$0.00020$  & $0.00455$ & $0.0252$ & $0.0729$ & $0.1510$\\

$^{217}_{\,\,\,89}$Ac &9.832 &$0.001126$ &$0.02741$&$0.1233$ &$0.2684$ &$0.4129$ \\

 \hline
\end{tabular}}
\caption{Average excitation energies evaluated using equation 
\eqref{excitation2}
in conjunction with \eqref{ericson}.}\label{table1}
\end{table}

\subsection{Perrone and Clayton approach}
The unique attempt in literature for the evaluation of the thermally enhanced 
$\alpha$ decay half-lives of nuclei in stellar environments can be found 
in \cite{clayton} by Perrone and Clayton. They evaluated the $\alpha$ decay 
rates of several nuclei formed in the s-process nucleosynthesis to find that 
the half-lives are greatly enhanced if one considered the 
contributions from thermally excited nuclei at elevated temperatures. 
The temperature-dependent half-life in \cite{clayton} was given by 
\begin{equation}
[t_{1/2}^{PC}(Z,A,T)]^{-1}=\int_{0}^{\infty}{\sum_J\frac{F(Z,A,E,J,T)\,
D(Z,A,E,J)dE}{t_{1/2}(Z,A,E,J)}}\label{clayton1}
\end{equation}
where $t_{1/2}(Z,A,E,J)$, the temperature-independent half-life for the 
decay of the parent nucleus to the daughter particle ground state is weighed 
by the nuclear density of states $D(Z,A,E,J)$ and the occupation probability 
$F(E,J,T)$, which for an excited nucleus with energy $E_i$ and spin $J_i$ is 
given by 
\begin{equation}
F(E_i,J_i,T)=\frac{(2J_i+1)e^{-E_i/k_B T}}
{\sum_i(2J_i+1)e^{-E_i/k_B T}}\approx\frac{(2J_i+1)e^{-E_i/k_B T}}{2J_0+1}
\end{equation}
with $J_0$ the spin of the ground state and $k_B$ the Boltzmann constant. 
The last approximation was justified by mentioning that the nuclear ground 
state dominates the sum over states for temperatures below 2 GK. 
The nuclear level density used was taken from \cite{gilbert}. 
The temperature-independent half-life in the denominator of equation \ref{clayton1} was 
calculated using the standard semiclassical approach for tunneling decay
where the penetration probability depends on the $Q$-value of $\alpha$ decay. 
The tunneling $\alpha$ is assumed to have an energy equal to the $Q$-value plus
the excitation energy (which is indeed the dummy variable of the integral in 
\ref{clayton1}). Since the penetration factor is evaluated within a simple 
model for the potential, the calculation of the temperature independent 
half-life appearing inside the integral in equation \ref{clayton1} 
essentially resembles the half-life as found in textbooks 
\cite{BookBeisser} but  
evaluated at an effective $Q$-value. 
With such a half-life as an input, the temperature dependent half-life 
was evaluated as an integral over energies from 0 to $\infty$. 

It is important to note that although the nuclear energy levels are discrete,
the half-life $t_{1/2}^{PC}(Z,A,T)$ defined by (\ref{clayton1}) in 
\cite{clayton} treats the nucleus as having a continuum of excited states
with the density of states given by $D(Z,A,E,J)$. The authors mention the 
need for explicitly taking the discrete levels into account if one wished 
to use the calculations in context with astrophysics. However, given the 
meagre data available in 1970, the authors found their approach appropriate 
for an initial survey of the problem. The present work takes into account 
this omission made by Perrone and Clayton due to the lack of data and finds
that even if the decrease in half-lives is not as spectacular as that in 
\cite{clayton}, it is surely significant and possibly 
relevant for nucleosynthesis calculations. 

\section{Excited nuclear levels, $Q$ values and shell closures}
\label{structure}
A naive expectation for the decay of thermally excited nuclei would be that 
for a nucleus which decays by $\alpha$ decay in its ground state, there must 
exist some excited levels which decay by emitting an $\alpha$ too.   
However, experimental results show that this is, in general, not true.  
A careful examination of the nuclear data tables reveals that 
the $\alpha$ decay occurs in heavy nuclei mostly in the ground state.
An excited parent nucleus often decays 
by emitting a photon ($\gamma$-decay). In fact, the excited nucleus 
undergoes several successive $\gamma$-decays before reaching its ground state. 
If this were true for all nuclei decaying by $\alpha$ decay, an undertaking as 
in the present work would not make much sense. 
However, based on a conjecture in an earlier work \cite{kelkar3}, we did 
find exceptions.

In \cite{kelkar3}, the authors noted an interesting phenomenon while 
performing a calculation of the tunneling times in $\alpha$ decay. 
The amount of time spent by an $\alpha$ in front of the barrier before 
tunneling (the so-called transmission dwell time), reaches a minimum 
at $N$ = 128 of the parent nucleus in the region from $N$ = 116 to 
$N$ = 132 which was investigated. 
$N$ = 128 of the parent however corresponds to $N$ = 126 of the daughter, 
implying that the $\alpha$ spends the least amount of time with the 
magic daughter. The latter essentially corresponded to the shortest half-life 
or the highest decay rate for a nucleus with $N$ = 128. 
Putting it differently, one can say that a parent nucleus 
decaying by emitting an $\alpha$ does it readily when the daughter happens to 
be at the shell closure of $N$ = 126. 
Motivated by this finding in \cite{kelkar3}, we examined the list of 
excited levels of nuclei with $N$ = 128 and with the possibility of an 
$\alpha$ decay in the ground state. Not much to our surprise, indeed, we 
found that the nuclei, $^{212}_{84}$Po, $^{214}_{86}$Rn, 
$^{215}_{87}$Fr, $^{216}_{88}$Ra and $^{217}_{89}$Ac had several 
experimentally observed $\alpha$ decays from excited levels. 
These nuclei are formed in the $r$-process nucleosynthesis and will be 
studied in the present work.  

The above phenomenon of a larger number of excited levels decaying by $\alpha$ 
decay should in principle happen at the other $N$ as well as $Z$ 
shell closures too. Inspecting 
the parent nuclei near $N$ or $Z$ = 84 we find that they do display 
some such excited states, but the effect is either not so pronounced 
or the data is scarce. In the range 
of the medium heavy nuclei, with daughters 
near the shell closure of $Z$ = 50, there are 
hardly any nuclei decaying by $\alpha$ decay and near $N$ = 50, none. 
We have an interesting case however at the lowest magic number of 2. 
$^8$Be decays 
to two $\alpha$'s, i.e., $^4$He nuclei and hence both the daughters in the 
decay have $N$ as well as $Z$ = 2. The number of excited levels which 
decay by $\alpha$ decay are 13 and one would expect a strong effect in 
the thermally enhanced decay rates. However, one does not observe a big 
change in the decay rate due to temperature since the spacing between levels 
is much larger than those in the heavy nuclei. For example, the first excited 
state in $^8$Be is around 3 MeV and $^{212}$Po has about 50 excited levels 
between 0 and 3 MeV. The high density of excited states in heavy nuclei is 
expected and was explained a long time ago by Bethe \cite{bethe1936}. 
\begin{figure}[ht]
\centering
\includegraphics[scale=0.5]{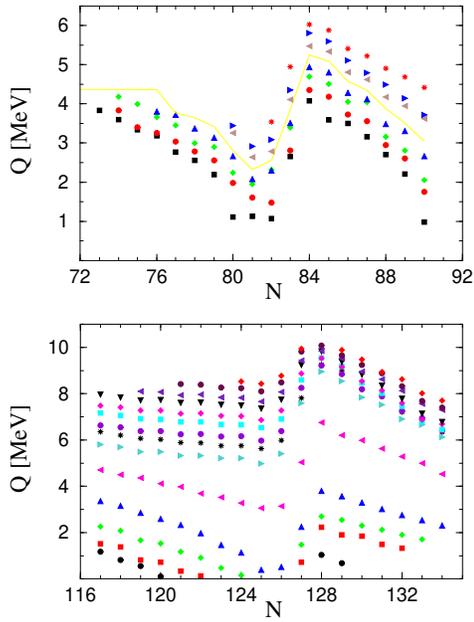}
\caption{Q values in $\alpha$ decay as a function of the 
neutron number of the parent nuclei. 
Similar symbols indicate nuclei with the same number of protons. 
Note the steep rise at
$N$ = 84 and $N$ = 128 corresponding to the shell closures of 
82 and 126 in the daughter nuclei.}\label{Qvaluesplot}
\end{figure}

The peculiar observations mentioned above are in fact a reflection of the 
behaviour of the $Q$-values in $\alpha$ decay as a function of the neutron 
and proton numbers, $N$ and $Z$ in the decaying nuclei. In Fig. 
\ref{Qvaluesplot} 
we see that close to the neutron numbers of $N^m$ + 2 
with $N^m$ being the magic numbers 82 and 126 
the $Q$ values rise sharply. Higher $Q$ values imply 
a large tunneling probability for the $\alpha$ and hence a bigger decay 
rate. Though not shown in the figure such a steep rise happens also as 
a function of the proton number, $Z$ for $Z^m$ + 2 with $Z^m$ being magic.
However, there are not many $\alpha$ decays in the vicinity of $Z$ = 52 and 
84.   
When the parent nucleus has a neutron number 
of $N^m$ + 2, we expect the decay probability to be large.  
The $Q$ value as such is a difference of the masses of nuclei and hence 
is a function of the binding energies of nuclei. 
The cluster preformation probability was 
shown in \cite{deng} to be directly proportional to the intrinsic energy 
of the cluster which in turn depends on the difference of binding 
energies of the nuclei involved. 
A larger clustering leads to a larger probability 
of populating excited states which decay by emitting an $\alpha$ 
\cite{buck5,astier,bairen}. The latter is due to the fact that increasing 
the energy increases the probability of tunneling of an already formed 
$\alpha$ cluster.    

Looking back at the heavy nuclei studied by Perrone and Clayton (for 
which data was most likely not available in 1970), namely, 
$^{144}$Nd, $^{148}$Sm, $^{150}$Sm, $^{152}$Sm, $^{158}$Dy, $^{174}$Hf and 
$^{176}$Hf, we do not find experimental evidence of 
excited levels decaying by $\alpha$ emission in any of
these nuclei. This is essentially due to the fact that the $Q$ values in these
decays are relatively small. 
These nuclides have long half-lives (of the order of 10$^{15}$ years) 
or are stable. From a potential barrier perspective, their probability of 
transmission is too small due to the fact that the Q-value is small.
One could consider the possibility of a $\gamma$ delayed 
$\alpha$ decay of a nucleus due to the thermally excited levels. However, the 
$\gamma$ decay half-lives of the excited levels are much smaller than the 
ground state $\alpha$ decay half-lives and the delay would be negligible. 

\section{Results and Discussions}
\label{results}
In view of the physics discussed in the previous section, we shall present 
calculations for the temperature dependence of half-lives of nuclei with 
the neutron number, $N$ = 128. The daughter nucleus in this case is at the 
shell closure of $N$ = 126. To be specific, we shall consider the 
following nuclei:
$_{\,\,\,84}^{212}$Po, $^{214}_{\,\,\,86}$Rn, $^{215}_{\,\,\,87}$Fr, 
 $^{216}_{\,\,\,88}$Ra and $^{217}_{\,\,\,89}$Ac. Each of these nuclei decays 
to a daughter nucleus with $N$ = 126. $_{\,\,\,84}^{212}$Po decays to the 
doubly magic nucleus $^{208}_{\,\,\,82}$Pb.
 
In the case of the statistical approach involving a sum over all excited 
states, we perform the calculation using the information provided in 
Table \ref{table2}. In the absence of data on the half-lives, we use a 
universal decay law (UDL) at a shifted $Q$-value, namely, $Q_i = Q + E_i$, 
where $E_i$ is the energy of the $i^{th}$ excited state of the parent nucleus.  
For the UDL calculation, we use the minimum allowed value of 
the orbital angular momentum
quantum number $l$ for each excited level decay and have 
listed it in Table \ref{table2}. Higher values of $l$ could be included,
however, the nature of the UDL is such that even if larger values of $l$
were taken into account, the final result would not change much.
Note that the difference between the $b$ and $d$ parameters of the second
and third term in Eq. (\ref{universallaw}),
respectively, is of two orders of magnitude
(the third term being the smallest). The ratio of these two terms grows
roughly linearly as a function of angular momentum $1/10 \times \sqrt{l(l+1)} \sim l/10$, therefore only very high values of angular momentum would make the third term comparable to the second. Those large momenta would happen for large values of $J_p$ and in such cases $l_{min}$ is also large providing a good
approximation to the half-life.
For the half-life using the UDL, we do not include any preformation factor, 
i.e., we assume it to be unity. 
The factor is usually calculated
phenomenologically from the ratio of the theoretical and experimental
half-lives. Performing such a calculation for each individual excited level
is a formidable task and out of the scope of the present work.
Besides, since the contribution of this factor will not vary exponentially,
it appeared to us a reasonable assumption to take it to be a constant. We
are interested in the relative decrease in half-lives at elevated temperatures
as compared to the terrestrial ones and expect that the error introduced
due to this omission is not large.

\begin{table}[ht]
\renewcommand{\arraystretch}{0.5}
\resizebox{\textwidth}{!}{
\centering
\begin{tabular}{|p{1.2cm}|p{1.3cm}|p{1cm}|p{1cm}|p{0.8cm}|p{1.8cm}|p{3.2cm}|}\hline
 Parent & $E_i[MeV]$ & $J^{\pi}_p$ &$J^{\pi}_d$& $l_{min}$&\% B.R ($\alpha$) & $t_{1/2}^{i}[s]$\\
\hline
 $_{84}^{212}$Po & 0      & 0+ &0+& 0 & 100   & $2.9\times 10^{-7}$ \\
               -  & 0.727 & 2+ &0+& 2 &0.033 & $1.42\times 10^{-11}$\\
               -  & 1.132 & 4+ &0+& 4 & 0.5   &$1.12\times 10^{-8}$ (UDL) \\
               -  & 1.249 & 0+ &0+& 0 & 100  & $1.52\times 10^{-10}$ (UDL)\\
               -  & 1.355 & 6+ &0+& 6 & 3     & $7.60\times 10^{-10}$\\
               -  & 1.476 & 8+ &0+& 8 & 3     & $1.46\times 10^{-8}$\\
               -  & 1.547 & 0+ &0+& 0 & 100   & $3.49\times 10^{-11}$ (UDL)\\
               -  & 1.578 & 0+ &0+& 0 & 100   & $3.01\times 10^{-11}$ (UDL)\\
               -  & 1.612 & 0+ &0+& 0 & 100   & $2.55\times 10^{-11}$ (UDL)\\
               -  & 1.657 & 0+ &0+& 0 & 100   & $2.06\times 10^{-11}$ (UDL)\\
               -  & 1.679 & 2+ &0+& 2 & 0.3   & $5.40\times 10^{-13}$\\
               -  & 1.800 & 0+ &0+& 0 &26    & $1.05\times 10^{-11}$ (UDL)\\
               -  & 1.805 & 2+ &0+& 2 &1.6   & $7.85\times 10^{-11}$ (UDL)\\
              -   & 2.930 & 18+&0+& 18 &99.93(96.83) & $45.1$\\
              -   & 2.930 & 18+&3-& 15 & 99.93(1) & $45.1$ \\  
              -   & 2.930 & 18+&5-& 13 & 99.93(2.05) & $45.1$ \\                   
 $_{86}^{214}$Rn  & 0     & 0+ &0+& 0 & 100   & $2.7\times 10^{-7}$\\
                - & 1.442 & 6+ &0+& 6 & 2$^{\dagger}$   & $6.9\times 10^{-8}$\\
                - & 1.625 & 8+ &0+& 8 & 4.1   & $6.5\times 10^{-9}$\\
 $_{87}^{215}$Fr  & 0     & 9/2-&9/2-& 0 & 100  & $8.6\times 10^{-8}$\\
                - & 0.835 & 13/2+ &9/2-& 3 &4.3 & $1.43\times 10^{-8}$ (UDL)\\
                - & 1.121 & 17/2- &9/2-& 4 &0.9& $8.07\times 10^{-9}$ (UDL)\\
                - & 1.149 & 15/2- &9/2-& 4 & 0.9& $7.04\times 10^{-9}$ (UDL)\\
                - & 1.440 & 19/2- &9/2-& 6 & 4.7 & $4.0\times 10^{-9}$\\
                - & 1.573 & 23/2- &9/2-& 8 & 4.1 & $3.5\times 10^{-9}$\\
$_{88}^{216}$Ra   & 0     & 0+    &0+& 0 & 100 & $1.82\times 10^{-7}$\\
                - & 1.164 & 4+    &0+& 4 & 0.23 &$1.61\times 10^{-8}$ (UDL)\\
                - & 1.507 & 6+    &0+& 6 &0.58 & $2.0\times 10^{-10}$\\
                - & 1.711 & 8+    &0+& 8 & 1.86 & $1.42\times 10^{-9}$\\
                - & 2.026 & 10+   &0+& 10 & 0.12 & $6.0\times 10^{-10}$\\
$_{89}^{217}$Ac   & 0     & 9/2-  &9/2-& 0 & 100  & $6.9\times 10^{-8}$\\
                - & 1.498 & 19/2- &9/2-& 6 & 0.46 & $9.81\times 10^{-9}$ (UDL)\\
                - & 1.528 & 21/2- &9/2-& 6 &0.46 & $1.00\times 10^{-8}$\\
                - & 2.012 & 29/2+ &9/2-& 11 & 4.51(4.1) & $7.40\times 10^{-7}$\\
                - & 2.012 & 29/2+ &7/2-& 11 &4.51(0.32) & $7.40\times 10^{-7}$\\
                - & 2.012 & 29/2+ &13/2+& 8 & 4.51(0.122) & $7.40\times 10^{-7}$\\
 \hline
\end{tabular}}
\\
${\dagger}$ since the listed value in the data tables is \%$\alpha >$0, we 
choose 2\% for an estimate 
\caption{Energy level, spin, parity, branching ratio and measured half-lives of levels which decay by alpha emission. If the experimental half-life of a level 
is not known, it is calculated using the UDL at an effective $Q$-value given 
by $Q + E_i$, where $E_i$ is the energy of the excited level. In such cases, 
even if the experimental branching ratio is known, it is not used but taken to 
be 100\% since the UDL per definition is formulated only for the alpha decay 
channel. $l_{min}$ is the minimum value of the orbital angular momentum quantum number, allowed by the selection rules.}\label{table2}
\end{table}

We provide two sets of results for the half-lives of nuclei with 
neutron number $N$ = 128 evaluated using two prescriptions. In the first set, 
Table \ref{table3}, we calculate half-lives using the double-folding 
approach and we compare them with the half-lives obtained by utilizing 
Eq. \eqref{universallaw}. We take $Q\rightarrow Q_{eff}$, with the 
$Q_{eff}$ approach of section \ref{Qeff-model}, in which the extra 
energy is taken as in Eq. \eqref{excitation2}. 
In our second set, Table \ref{table4}, we use the statistical method where 
the decay constant and, in turn the half-life, are given by 
Eq. \eqref{stathalflife}. We compare two approaches in Table \ref{table4}: 
one in which all available listed levels are included and another one in 
which only levels experimentally found to decay by alpha emission are used. 
For those levels (in both approaches) with unknown experimental half-lives, 
the UDL (Eq. \eqref{universallaw}) is invoked to estimate $t_{1/2}$ of those 
levels. 
The results in Table \ref{table3} ensure that the estimate obtained 
from Eq. \eqref{universallaw} and used as 
an input in the statistical approach (for the experimentally 
unknown half-lives) is reasonable. 
In Table \ref{table3}, a comparison of the half-lives using the UDL expression 
and the double folding model is given for a range of temperatures from 
0 to 2.4 GK. The half-life at a given temperature is determined using an effective $Q$ value given by Eq.\eqref{effectiveQ}, namely, $Q_{eff}=Q + \bar{\epsilon}(A,Z,T)$ where $\bar{\epsilon}(A,Z,T)$ is the average excitation energy at a given temperature. The latter is evaluated considering all excited levels of a given 
nucleus. 
For the sake of comparison we shall choose the preformation factor, 
$P_{\alpha}$ = 1 in Eq. (\ref{width}).
 
\begin{table}[ht]
\resizebox{\textwidth}{!}{
\begin{tabular}{ |p{2.2cm}|p{1cm}|p{2.1cm}|p{1.9cm}|p{1.9cm}|p{1.9cm}| p{1.9cm}|p{1.9cm}|}
 \hline
 \multicolumn{8}{|c|}{$t_{1/2}(T)$[s]} \\
 \hline
&$Q$&$0 GK$&$0.8GK$&$1.2GK$&$1.6GK$& $2GK$&$2.4GK$\\
 \hline

$^{212}Po$ [UDL] &8.954 &$1.572\times 10^{-7}$ &$1.57\times 10^{-7}$  & $1.54\times 10^{-7}$ & $1.36\times 10^{-7}$ & $9.57\times 10^{-8}$ & $5.69\times 10^{-8}$\\
& & &(0.06)& (2.09) & (13.2) & (39.1) & (70.1)\\
$^{212}Po$ [DFM] & &$6.49\times 10^{-8}$ &$6.48\times 10^{-8}$  & $6.36\times 10^{-8}$ & $6.45\times 10^{-8}$ & $4.12\times 10^{-8}$ & $2.16\times 10^{-8}$\\
& & &(0.15)& (1.92) & (12.1) & (36.4) & (66.7)\\

$^{214}Rn$ [UDL] &9.208 &$2.07\times 10^{-7}$ &$2.06\times 10^{-7}$  & $2.01\times 10^{-7}$ & $1.77\times 10^{-7}$ & $1.31\times 10^{-7}$ & $7.49\times 10^{-8}$\\
& & &(0.09)& (2.60) & (14.0) & (36.4) & (61.6)\\
$^{214}Rn$ [DFM] & &$7.92\times 10^{-8}$ &$7.91\times 10^{-8}$  & $7.73\times 10^{-8}$ & $6.89\times 10^{-8}$ & $5.23\times 10^{-8}$ & $3.31\times 10^{-8}$\\
& & &(0.08)& (2.39) & (12.98) & (33.9) & (58.2)\\

$^{215}Fr$ [UDL] &9.540 &$7.11\times 10^{-8}$ &$7.07\times 10^{-8}$  & $6.07\times 10^{-8}$ & $3.28\times 10^{-8}$ & $1.28\times 10^{-8}$ & $5.32\times 10^{-9}$\\
& & &(0.55)& (14.6) & (53.8) & (81.9) & (92.5)\\
$^{215}Fr$ [DFM] & &$2.92\times 10^{-8}$ &$2.90\times 10^{-8}$  & $2.53\times 10^{-8}$ & $1.45\times 10^{-8}$ & $6.19\times 10^{-9}$ & $2.79\times 10^{-9}$\\
& & &(0.50)& (13.4) & (50.4) & (78.8) & (90.5)\\

 $^{216}Ra$ [UDL] &9.526 &$1.862\times 10^{-7}$ &$1.86\times 10^{-7}$  & $1.81\times 10^{-7}$ & $1.60\times 10^{-7}$ & $1.21\times 10^{-7}$ & $7.76\times 10^{-8}$\\
 & & &(0.93)& (2.63) & (13.7) & (34.6) & (58.3)\\
 $^{216}Ra$ [DFM] & &$6.86\times 10^{-8}$ &$6.86\times 10^{-8}$  & $6.70\times 10^{-8}$ & $6.00\times 10^{-8}$ & $4.66\times 10^{-8}$ & $3.1\times 10^{-8}$\\
& & &(0.09)& (2.40) & (12.6) & (32.1) & (54.9)\\
    
$^{217}Ac$ [UDL] &9.832 &$7.67\times 10^{-8}$ &$7.61\times 10^{-8}$ &$6.56\times 10^{-8}$&$3.84\times 10^{-8}$ &$1.73\times 10^{-8}$ &$7.96\times 10^{-9}$ \\
& & &(0.63)& (14.3) & (49.8) & (77.4) & (89.6)\\
$^{217}Ac$ [DFM] & &$2.98\times 10^{-8}$ &$2.96\times 10^{-8}$  & $2.59\times 10^{-8}$ & $1.59\times 10^{-8}$ & $7.75\times 10^{-9}$ & $3.89\times 10^{-9}$\\
& & &(0.56)& (13.0) & (46.5) & (74.0) & (87.1)\\           
 \hline
\end{tabular}}
\caption{Half-lives within the effective $Q$-value approach with the first row 
displaying the half-lives using the UDL at, $Q_{eff} = Q + \bar{\epsilon}$ (with $\bar{\epsilon}$ given by \eqref{excitation2} and listed in table \ref{table1}). 
The second row displays the half-lives obtained by using DFM and $Q_{eff}$. 
Quantities in brackets show the percentage decrease in the half-life due 
to temperature.}
\label{table3}
\end{table}
From Table \ref{table3}, we see that the increase in temperature in general decreases the half-lives with the decrease being at the most an order of magnitude from $T$ = 0 to 2.4 GK. Though the half-lives evaluated using the universal decay law are not exactly the same as those in the more realistic 
double folding model, the percentage decrease in both cases is roughly the 
same. The percentage decrease is calculated as  
\begin{equation}
pd=\frac{t_{1/2}^{T=0}-t_{1/2}^T}{t_{1/2}^{T=0}}\times 100
\end{equation}
This fact allows us with some reliability to use the UDL 
given by \eqref{universallaw} in the statistical approach 
for the calculation of the missing half-lives 
in the available data for excited levels. 

The temperature-dependent half-lives for several isotopes using the 
statistical approach are displayed in Table \ref{table4}.
As the temperature increases the half-lives are reduced and the reduction 
is larger than found in the effective $Q$-value approach.  
The calculations are done using the experimentally listed half-lives. 
For these 
particular isotopes, several excited levels have been observed to decay 
by alpha decay, however, the half-lives of many of these excited states 
have not been measured. For the cases with no experimental information, 
we use the UDL to evaluate the half-lives to be used in \eqref{Iliadis}.  
The two rows labelled [A] and [B] in the table display the calculations 
including half-lives of all listed levels in \eqref{Iliadis} and only those 
which decay by $\alpha$ decay, respectively. 
Particularly interesting in this table is the case of $^{212}$Po. This nucleus 
decays to the doubly magic nucleus $^{208}$Pb and an $\alpha$ which could 
possibly be the reason (as argued in earlier sections) that $^{212}$Po has 
many more excited levels which decay by emitting an $\alpha$ as compared to 
the other nuclei in the table. 

As mentioned above, the calculation labeled as [A] includes decay from 
all parent states. The results are, in most cases, very similar to those 
when only the levels experimentally known to decay by emitting an alpha 
(labeled as [B]) are used. This is a good indication that a general simpler 
formulation would be appropiate and would permit an extension to 
more nuclei, thus facilitating nucleosynthesis calculations.
\begin{table}

\resizebox{\textwidth}{!}{
\begin{tabular}{ |p{2.2cm}|p{1cm}|p{1.9cm}|p{2.2cm}|p{2.2cm}|p{2.3cm}| p{2.3cm}|p{2.3cm}|}
 \hline
 \multicolumn{8}{|c|}{$t_{1/2}(T)$[s]} \\
 \hline
Isotope&$Q$&$0$ GK&$0.8$ GK&$1.2$ GK&$1.6$ GK& $2GK$&$2.4$ GK\\
 \hline
 
$^{212}Po [A]$ &8.954 &$2.99\times 10^{-7}$ &$2.913\times 10^{-7}$  & $3.502\times 10^{-9}$ & $6.27\times 10^{-11}$ & $5.86\times 10^{-12}$ & $1.31\times 10^{-12}$\\

$^{212}Po [B]$ & &$2.99\times 10^{-7}$ &$2.916\times 10^{-7}$  & $3.611\times 10^{-9}$ & $6.45\times 10^{-11}$ & $6.01\times 10^{-12}$ & $1.32\times 10^{-12}$\\

\hline

$^{214}Rn [A]$ &9.208 &$2.7\times 10^{-7}$ &$2.59\times 10^{-7}$  & $1.22\times 10^{-7}$ & $3.36\times 10^{-8}$ & $7.67\times 10^{-9}$ & $1.04\times 10^{-9}$\\

$^{214}Rn [B]$ & &$2.7\times 10^{-7}$ &$2.69\times 10^{-7}$  & $2.68\times 10^{-7}$ & $2.67\times 10^{-7}$ & $2.51\times 10^{-7}$ & $2.0\times 10^{-7}$\\

\hline

$^{215}Fr [A]$ &9.540 &$8.6\times 10^{-8}$ &$8.53\times 10^{-8}$  & $7.09\times 10^{-8}$ & $3.68\times 10^{-8}$ & $1.22\times 10^{-8}$ & $2.60\times 10^{-9}$\\

$^{215}Fr [B]$ & &$8.6\times 10^{-8}$ &$8.59\times 10^{-8}$  & $8.58\times 10^{-8}$ & $8.44\times 10^{-8}$ &$8.03\times 10^{-8}$ & $7.36\times 10^{-8}$ \\

\hline

$^{216}Ra [A]$ &9.526 &$1.82\times 10^{-7}$ &$1.79\times 10^{-7}$  & $1.29\times 10^{-7}$ & $4.39\times 10^{-8}$ & $9.88\times 10^{-9}$ & $1.4\times 10^{-9}$\\

$^{216}Ra [B]$ & &$1.82\times 10^{-7}$ &$1.81\times 10^{-7}$  & $1.69\times 10^{-7}$ & $7.68\times 10^{-8}$ & $1.99\times 10^{-8}$ & $6.52\times 10^{-9}$\\

\hline

$^{217}Ac [A]$ &9.832 &$6.9\times 10^{-8}$ &$6.85\times 10^{-8}$  & $5.87\times 10^{-8}$ & $3.19\times 10^{-8}$ & $1.20\times 10^{-8}$ & $4.41\times 10^{-9}$\\

$^{217}Ac [B]$ & &$6.9\times 10^{-8}$ &$6.89\times 10^{-8}$  & $6.88\times 10^{-8}$ & $6.87\times 10^{-8}$ & $6.86\times 10^{-8}$ & $6.82\times 10^{-8}$\\
           
 \hline
\end{tabular}}
\caption{Alpha decay half-lives at different temperatures evaluated using 
\eqref{Iliadis} within the statistical approach. The half-lives in the 
rows marked [A] are evaluated using \eqref{Iliadis}, including the 
entire set of tabulated levels. Rows marked [B] take into account only those 
levels which are experimentally known to decay by the emission of an $\alpha$.}\label{table4}
\end{table}


A small note on the comparison of the two approaches, namely, the effective $Q$-value approach and the
statistical approach is in order here. With the aim of providing temperature dependent half-lives for nucleosynthesis
applications, we began by formulating an approach where the effective Q-value would enter in the UDL with the advantage
of avoiding the task of performing calculations of half-lives for many individual levels populated at very high temperatures, the
use of a density level model; thus, making the proposed effective $Q$-value model feasible in a network calculation.
However, this advantage comes at the expense of missing information about the variation in the half-lives of
different levels as well as the population probability of the excited states. Performing an average over the excitation
energies and calculating for just one effective $Q$-value is equivalent to considering decay from one excited level
which occurs at an effective excitation energy. A more realistic description is provided in the statistical approach,
which would be model-independent as long as the half-lives of the excited levels and the branching ratios for alpha
decay are known. Our introduction of the UDL for the unknown half-lives gives a pathway to extend the calculations
to a larger number of alpha emitters. In this work, we use the known UDL from literature and evaluate the half-life
of an unknown level with energy $E_i$, by replacing the $Q$-value in the UDL by $Q + E_i$. However, formulating a UDL
for excited levels using all available data on the alpha decay of excited parent and daughter nuclei is a task which
we plan for the future. 

Finally, in passing we mention that there also exists the possibility that the 
system can transit from the excited state to the lower state and then
decay by alpha. However, given the exponential nature of the population probability,
namely, $p_i \propto \exp{(- E_i/kT)}$ (with $E_i$ being the energy of the
excited state), at a given temperature,
the likelihood of a higher level being populated and
decaying from a lower level to which it transits will surely be smaller
than the lower level itself getting populated and decaying by emitting
an alpha. Furthermore, in the cases where experimental information exists,
the branching fraction, BR$_{ij}$, accounts for the effect.

Alpha decay plays a role, in competition with beta decay and fission, 
in powering and shaping the light curves of kilonovae 
\cite{Barnes:2013wka,Barnes:2016umi,Barnes:2020nfi}. It is customary in 
nucleosynthesis calculations to consider alpha emission only from the 
ground state, i.e at zero temperature. In view of the reduction of the 
$\alpha$-decay half-lives in hot 
environments found above, it seems appropriate to replace 
these inputs by temperature dependent ones. Such a detailed calculation, 
though necessary is out of scope of the present work. 

\section{Summary}
\label{conclusions}
In this work we have explored the role of temperature in alpha emission 
from nuclear excited states. We used a statistical approach and proposed a 
model that can potentially be extended for several nuclei. 
We particularly focused on nuclei that can be produced in $r$-process, 
motivated by the impact that alpha decay has on the heating of light curves 
of kilonovae. Thermally enhanced alpha decay rates were calculated for nuclei
with the neutron number $N$ = 128 decaying to a daughter nucleus at the shell 
closure with $N$ = 126. The latter choice was made due to the occurrence of 
more excited levels decaying by $\alpha$ emission as compared to other nuclei.
The calculation performed within the statistical approach is in principle 
model independent. It requires the experimental input of the energies, 
spins and half-lives (as well as their branching fractions for $\alpha$ 
decay) of the excited levels. However, sometimes the 
experimental data are incomplete (e.g. even if it
is known that an excited level decays via $\alpha$ decay, its 
half-life is not known). 
In such a case we supply the missing information by calculating the 
half-life using a universal decay law (UDL) for the half-life. 
The latter introduces some uncertainty in the results, however, we do not 
expect the results to change drastically since the  
temperature dependence of the half-lives using the UDL is in good agreement
with the predictions of the more elaborate double folding model for tunneling 
decay. 

We found that temperatures of the order of GK can increase the half-lives of 
the nuclides studied here by at least a factor of ten. Particularly, 
for the case of $^{212}$Po and depending on the model, 
the change can be by orders of magnitude. 

\section{acknowledgments}
O.L.C. thanks Jose Trujillo and Fernando Montes for interesting discussions and acknowledges the support of the Natural Science and Engineering Research Council of Canada (NSERC).

\end{document}